\documentstyle[twocolumn,aps]{revtex}

\begin{document}
\draft

\twocolumn[\hsize\textwidth\columnwidth\hsize\csname %
@twocolumnfalse\endcsname

\title{Critical Dynamics of a Vortex Loop Model
 for the Superconducting Transition}
\author{Vivek Aji and Nigel Goldenfeld}

\address{Department of
Physics, University of Illinois at Urbana-Champaign
\\1110 West Green Street\\
Urbana, IL, 61801-3080}

\date{\today}

\maketitle

\begin{abstract}

We calculate analytically the dynamic critical exponent $z_{MC}$
measured in Monte Carlo simulations for a vortex loop model of the
superconducting transition, and account for the simulation results. In
the weak screening limit, where magnetic fluctuations are neglected,
the dynamic exponent is found to be $z_{MC} = 3/2$. In the perfect
screening limit, $z_{MC} = 5/2$.  We relate $z_{MC}$ to the actual
value of $z$ observable in experiments and find that $z \sim 2$,
consistent with some experimental results.
\end{abstract}

\vspace{0.2in}

]

The discovery of the short coherence length cuprate superconductors has
allowed heretofore inaccessible fluctuation effects in superconductors
to be probed. Beginning with the penetration depth measurements of
Kamal et al. \cite{Kamal}, and including measurements of magnetic
susceptibility \cite{LBH,SAL1}, resistivity \cite{SAL1,HWS} and
specific heat \cite{OVR2}, static and dynamic fluctuation effects have
been convincingly observed and accurately quantified.  These
measurements are consistent with the theory of a strongly type-II
superconductor, with a weak coupling of the order parameter to the
electromagnetic field, described by the $3$D XY model coupled to a
gauge field\cite{FFH}.

The dynamic critical exponent, $z$, characterizes the relaxation to
equilibrium of fluctuations in the critical regime of systems
exhibiting a second order phase transition \cite{HH,NGbook}. In
particular it relates the time scale of relaxation, $\tau$, to a
relevant length scale, $x$: $\tau \sim x^{z}$. For infinite systems $x$
is the correlation length, $\xi$. Near the critical point, the
correlation length diverges and the relaxation time tends to infinity,
a phenomenon known as critical slowing down. In finite size scaling
studies, $x$ is identified as the system size $L$.

The dynamic critical exponent, obtained from the measurement of
longitudinal dc-resistivity for YBCO is $z=1.5 \pm 0.1$ in finite but
small magnetic fields \cite{JTK}. Similar results were reported for the
zero-field DC conductivity \cite{HER,PDR}. Frequency dependent
microwave conductivity experiments yield $z \sim 2.3-3.0$ \cite{JCB}.
On reanalysis it was found that the data were consistent with $z \sim
2$ provided one neglected the region close to $T_{c}$ \cite{DRS}.
Moloni et al. obtained $z=1.25 \pm 0.05$ at low magnetic fields
\cite{MOL}, but a later, more complicated analysis by these authors
gave $z=2.3 \pm 0.2$.  More recently, DC conductivity measurements on
single crystal BSCCO samples were interpreted to give evidence for
$z\sim 2$ \cite{HAN}.  In summary, experiments do not yet yield a
consistent picture of the critical dynamics.

If the dynamic exponent were indeed $z\sim 1.5$ , then this would be
surprising.  Precisely this value is obtained for the superfluid
transition in He$^4$ where the combination of second sound (a
propagating mode, therefore $z=1$) and order parameter dynamics
(diffusive, therefore $z\sim 2$) lead to $z=3/2$ (model E dynamics)
\cite{HH}.  In YBCO, however, the combination of a momentum sink
arising from the lattice, and the Coulomb interaction destroying the
longitudinal current fluctuations should lead to pure order parameter
dynamics and a prediction that $z\sim 2$ (model A dynamics). It is of
course possible that some other mechanism can yield $z\sim 1.5$.

To shed light on these issues the critical dynamics was investigated
numerically by performing a Monte Carlo calculation of $z$ for the
$3$-dimensional XY model, in the vortex representation (the so-called
Villain model \cite{VIL}), with and without magnetic screening
\cite{LID}. The spin wave degrees were replaced by discrete vortex
variables and the dynamics imposed was dissipative. The dynamic
exponent estimated through a scaling analysis of the resistivity
calculated within linear response will be denoted by $z_{MC}$.
Surprisingly enough the exponent was found to be $z_{MC}\sim 1.5$ when
the interaction was unscreened while $z_{MC}\sim 2.7$ in the presence
of screening. Not only does the value $z_{MC}\sim 1.5$ agree with
previous results obtained by performing a similar analysis on the
London Lattice Model (LLM) \cite{JEN} but also with the value of $z$
reported in some of the experiments cited above.  The observations in
the computer simulations are surprising because there are no collective
modes in the Villain model so that the dynamics would be expected to be
purely diffusive, with $z_{MC}\sim 2$.  Nevertheless, and contrary to
expectation, here too the system seems to support model E dynamics.
Other extensive simulation studies report values of $z_{MC} \sim 1.5$
and $z_{MC}\sim 2$ depending upon the boundary conditions\cite{MIN}.

The purpose of this Letter is to calculate analytically the dynamic
exponent for the Villain model. The equation of motion, corresponding
to the Monte Carlo steps implemented in the numerical computation, is
derived and analyzed near equilibrium. A scaling analysis is used to
extract $z_{MC}$.  We are able to explain the simulation results in
both strong and weak screening limits.  We show also that the
simulation results cannot be interpreted as providing evidence in
support of the $z\sim 1.5$ result found in some experiments, because
they do not measure the true dynamic critical exponent: $z_{MC}\neq z$.
We show how to relate $z_{MC}$ and $z$, and find that the result
$z_{MC}\sim 1.5$ is in fact an artifact of taking the thermodynamic
limit and the range of vortex interactions to infinity limit in the
wrong order. The correct physical prediction from the simulation is
$z\sim 2$ for any finite range of interaction, consistent with some
observations.

{\it The Villain model:-} Consider the XY model with a fluctuating
vector potential $\vec{a}$ represented as lattice gauge theory link
variables $a_{ij}\equiv a_i - a_j$:

\begin{equation}
H = -J\sum_{\left<i,j\right>}
\cos(\phi_{i}-\phi_{j}-\lambda_{0}^{-1}a_{ij}) + {1 \over
2}\sum_{\Box}[\vec{\nabla} \times \vec{a}]^{2}
\end{equation}

{\noindent}where $J$ is the coupling constant, $\lambda_{0}$ is the
screening length, $\phi_{i}$ is the phase of the condensate on site $i$
of a simple cubic lattice of size $N = L^{3}$ with periodic boundary
conditions. The first sum is taken over nearest neighbors, while the
second is over plaquettes of the lattice. The lattice spacing has been
set to unity.   The fluctuating gauge
potential $a_{ij}$ satisfies the constraint that at each site $i$, the
discrete divergence vanishes: $[\vec{\nabla}\cdot \vec{a}]_{i}=0$. The
phase degrees of freedom can be replaced by vortices by introducing the
periodic Villain function to replace the cosines. Standard
manipulations \cite{KNT} lead to the dual Hamiltonian:

\begin{equation}
H_{V} = {1 \over 2}\sum_{i,j}\vec{n_{i}}\cdot
\vec{n_{j}}G_{ij}[\lambda_{0}]
\end{equation}

{\noindent}where the $\vec{n_{i}}$'s are vortex variables that reside on
the links of the dual lattice and $G_{ij}$ is the screened lattice
Green's function,

\begin{equation}
G_{ij}[\lambda_{0}] = J{(2\pi)^{2} \over
L^{3}}\sum_{\vec{k}}{\exp[i\vec{k} \cdot (\vec{r_{i}} - \vec{r_{j}})]
\over {2\sum_{m}^{3}[1 - \cos(k_{m})] + \lambda_{0}^{-2}}}
\end{equation}

{\noindent}The two limits that are considered in the simulations are
the long range case, $\lambda_{0} \rightarrow \infty$, and the short
range case, $\lambda_{0} \rightarrow 0$. Actually the simulations were
performed by setting $\lambda_{0} = 0$ and $\lambda_{0} = \infty$ in (3).
The distinction between the limit and the actual simulations will turn
out to be significant. In both cases the local
constraint of no monopoles, $[\vec{\nabla} \cdot \vec{n}]_{i} = 0$, is
imposed. Each Monte Carlo move consists of trying to create a closed
vortex loop around a plaquette. The trial state is accepted or rejected
according to the heat bath algorithm with probability $1/[1 +
\exp(\beta\Delta E)]$ where $\Delta E$ is the change in energy and
$\beta = 1/k_BT$, with $k_B$ being Boltzmann's constant. Each time a
vortex loop is formed it generates a voltage pulse, $\Delta Q = \pm 1$,
perpendicular to its plane, the sign depending on the orientation. This
voltage fluctuation gives rise to an electrical resistance, $R$, which
can be analyzed within linear response theory. A point that will be
important to note here is that $R$ depends on the average change in the
total number of loops pointing in a given direction at each time step.
The unit of time is normalized so that, on average, an attempt has been
made to create or destroy one loop per plaquette.

{\it Dipole gas description:-} It is known that near $T_{c}$ the static
properties are dominated by the proliferation of vortex loops of unit
strength, i.e., it is energetically unfavorable to create vortex loops
of greater strength at each plaquette. The interaction between these
vortex loops is spherically symmetric and so is the state in thermal equilibrium.
As it stands, the computations above have been performed on what is known as
the low temperature Villain model and the critical point is obtained by
looking at the intersection of the low and high temperature Villain
models (for details see ref. \cite{KNT}). The physics described here is
that of an interacting gas of dipoles, $\vec {d}$. In the long range
case they interact via the standard Coulomb term which falls off as
$1/r^{3}$; note that these dipoles interact antiferromagnetically, and
are not current loops, which interact via the standard ferromagnetic
interaction.

For our analysis we shall consider a cubic lattice, $L^{3}$, on whose
vertices reside the loop variables, $\vec{l_{i}}$. In terms of the vortex
variables $\vec{n_{i}} = \vec{\nabla} \times \vec{l_{i}}$, as can be
seen by writing out the components. The three components are each
either $\pm 1$ or $0$, corresponding to a clockwise, anticlockwise or
absence of a vortex loop along the three principle directions, $x,y$ or
$z$. The corresponding probabilities on site $i$ at time step $s$ are
given by $P^{\alpha}_{is}[1], P^{\alpha}_{is}[-1]$ and
$P^{\alpha}_{is}[0]$, where $\alpha$ is a coordinate label. The quantity
computed in the simulations is the total number of loops, $N_{s}^{\alpha}$
pointing along a given direction $\alpha$ at time step $s$.

\begin{equation}
N_{s+1}^{\alpha} = \sum_{i}(P_{is+1}^{\alpha}[1] - P_{is+1}^{\alpha}[-1])
\end{equation}

To study the behavior of $N_{s}^{\alpha}$, we follow the
standard procedure of writing out the master equation for the time
development of the probabilities and evaluating (4) \cite{ISI}. As previously
indicated, the equilibrium state is spherically symmetric. That is, on average,
$\Delta E_{is}^{\alpha}$,
the change in energy on adding a unit loop on site $i$ at time $s$, is
zero. This implies that transition probabilities for creating and annihilating
a vortex loop are equal. The heat bath algorithm ensures that the conditions of
detailed balance are satisfied. Furthermore, at $T_{c}$, the restriction to
unit loops per plaquette results in $P^{\alpha}_{is}[0] = P^{\alpha}_{is}[1] =
P^{\alpha}_{is}[-1]$ in equilibrium.
Since we are interested in small deviations from equilibrium, we impose a
uniform perturbation, $\delta l^{\alpha}$ per site and see how it
relaxes back to equilibrium. This implies $\delta N^\alpha \sim L^{3}\delta
l^\alpha$. To leading order the equation of motion reads

\begin{equation}
{d\delta N^{\alpha} \over dt} = -{2 \over
3}\beta\sum_{i}a_{i}^{\alpha}\left({\partial \Delta E_{i}^{\alpha} \over
{\partial l^{\alpha}}}\right)_{0}\delta l^{\alpha}
\end{equation}
\noindent where the subscript $0$ denotes equilibrium, and
$a_{i}^{\alpha}$ is the transition probability in equilibrium for
creating the dipole loops.

{\it Scaling analysis:-} Equation (5) is the basis for the scaling
analysis that follows. The only relevant length scales are the system
size, $L$, and the correlation length, $\xi$. $a_{i}^{\alpha}$ is an
equilibrium microscopic transition probability which remains finite at
the critical point while $\beta\sum_{i}\Delta E_{i}^{\alpha}$ is
dimensionless and scales as $(L/\xi)^{3}$ away from $T_{c}$ for finite
systems. This follows because by definition, thermodynamic additivity
occurs on a scale beyond the correlation length. While the free energy
is extensive for all temperatures, at $T_{c}$, $\xi \sim L$ and
$\beta\sum_{i}\Delta E_{i}^{\alpha}$ is independent of $L$. Thus the
characteristic time scale of relaxation of the perturbation, $\tau$,
scales as
\begin{equation}
\tau \sim \frac{\xi^{3}[l]}{a_i^{\alpha}}
\end{equation}

\noindent where $[l]$ is the scaling
dimension of the field $l$.

For the long range case the binding energy is given by,

\begin{equation}
\beta H = -\beta\sum_{i,j}{\vec{d_{i}}\cdot \vec{d_{j}} -
3(\vec{d_{i}}\cdot {\hat{r}_{ij}})(\vec{d_{j}}\cdot
{\hat{r}_{ij}}) \over
{r^{3}}}
\end{equation}

{\noindent}where $\vec{d_{i}} = \mu \vec{l_{i}}$, $\mu$ is the dipole
strength of a unit loop around a plaquette, $r = |\vec{r}_{ij}|$, where
$\vec{r}_{ij} = \vec{r_{i}} - \vec{r_{j}}$ and ${\hat{r}_{ij}}$ is the
unit vector along $\vec{r}_{ij}$. If $\vec{l_{i}}$ were dimensionless,
then the energy of the system would not be extensive. To evaluate the
dimension of $\vec{l}$ note that $L^{6}[l]^{2}/\xi^{3} \sim
(L/\xi)^{3}$ as required by the extensivity of the free energy. Thus
$[l] \sim L^{-3/2}$ and $\tau \sim \xi^{3}L^{-3/2}$. The dynamic
exponent at $T_{c}$, where $\xi = L$, in this case is $z_{MC} = 3/2$,
which is consistent with the computer simulation results.

For the short range case the binding energy is given by

\begin{equation}
{\beta H} = {\beta \sum_{i}\vec{n_{i}} \cdot \vec{n_{i}}}
= {\beta \sum_{i}(\vec{\nabla} \times
\vec{d_{i}})\cdot (\vec{\nabla} \times \vec{d_{i}})}
\end{equation}

{\noindent}Requiring extensivity, i.e. $[l]^{2}\xi^{-2}L^{3} \sim
(L/\xi)^{3}$, yields $[l] \sim \xi^{-1/2}$. From (6) we get $\tau \sim
\xi^{5/2}$ which at $T_{c}$ scales as $L^{5/2}$. The dynamic exponent
is $z_{MC}=5/2$, which is consistent with the computer simulation
results \cite{LID}.

{\it Critical dynamics of the dipole gas model:-} We will now derive
the governing stochastic partial differential equation that describes
the long wavelength critical dynamics of the superconductor. Our
strategy will be to first derive the continuum limit of the Hamiltonian
(2), and then impose relaxational dynamics. We will find that the
results for $z$ are not the same as our results for $z_{MC}$. This is
because the Monte Carlo time step does not correspond to the physical
time step. This is explained below. Let us first look at the continuum
limit of the short range case. Reintroducing the coupling constants and
the lattice spacing, $a$, we write the Hamiltonian $H_{V}$ for the
vortex variables as

\begin{equation}
H_{V} =   J \left(2\pi{\lambda_{0} \over
a}\right)^{2}\sum_{i}(\vec{\nabla} \times \vec{l_{i}})\cdot (\vec{\nabla}
\times \vec{l_{i}})
\end{equation}

{\noindent} Converting the sum to an integral,

\begin{equation}
H_{V} = - (J/a^{3}) \left(2\pi{\lambda_{0} \over a}\right)^{2}\int d{\vec
{r}}\,(\vec{\nabla} \times \vec{l}(\vec{r}))\cdot (\vec{\nabla} \times
\vec{l}(\vec{r}))
\end{equation}

{\noindent}In the limit $a \rightarrow 0$, $Ja^{-3} \rightarrow
\widetilde{J}$ and $\lambda_{0}/a \rightarrow \widetilde{\lambda_{0}}$.
Relaxational dynamics is governed by the time-dependent Ginzburg-Landau
equation (TDGL), which in this case is

\begin{equation}
{\partial\vec{l} \over \partial t} = \Gamma
\widetilde{J}(2\pi{\widetilde{\lambda_{0}}})^{2}(\nabla^{2}\vec{l} -
\vec{\nabla}(\vec{\nabla}\cdot \vec{l})) + \vec{\eta}
\end{equation}

{\noindent}where $\eta$ is a white noise, satisfying the
fluctuation dissipation theorem with
$\left<\eta^{\alpha}(\vec{r})\right> = 0$ and
$\left<\eta^{\alpha}(\vec{r} ')\eta^{\beta}(\vec{r})\right> = 2\Gamma
k_{B}T\delta_{\alpha \beta}\delta(\vec{r} ' - \vec{r})$. The TDGL
equation is similar to the diffusion equation and is expected to yield
a dynamic exponent of $z=2$, in mean field theory, with small
corrections due to fluctuations.  The linearity of the TDGL in this
case reflects the fact that only unit vortices are considered in the
analysis.

In the long range case, taking the continuum limit, we obtain

\begin{eqnarray}
H_{V} = \widetilde{J}(2\pi)^{2}\int d{\vec{r}'} d{\vec{r}} &\,&
{(\vec{\nabla} \times \vec{l}(\vec{r}'))\cdot (\vec{\nabla} \times
\vec{l}(\vec{r})) \over {|\vec{r}' - \vec{r}|}}\\ \nonumber &\times&
\exp{[-|\vec{r} - \vec{r}'|/\lambda_{0}]}
\end{eqnarray}

{\noindent}where the infinite self energy has been subtracted, and the
screening length $\lambda_0$ is taken to be finite. To relate this to
the dipole-dipole interaction used in our analytic model of the
simulations, consider a cubic lattice, as before, on whose vertices sit
variables $\vec{d_{i}}$, and take $\lambda_{0} = \infty$. Replacing
$2\pi\sqrt{\widetilde{J}}\vec{l(\vec{r})} =
\sum_{i}\vec{d_{i}}\delta(\vec{r} - \vec{r_{i}})$ one can perform the
integrals over $\vec{r}$ and $\vec{r'}$, to recover the expression in
(7). The actual TDGL equation for the long range case reads

\begin{eqnarray}
{\partial\vec{l} (\vec{r}) \over \partial t} = \Gamma
\widetilde{J}(2\pi)^{2}\int d{\vec{r}'} &\,&{\nabla^{2}\vec{l}(\vec{r}') -
\vec{\nabla}(\vec{\nabla}\cdot\vec{l}(\vec{r}'))  \over {|\vec{r}' -
\vec{r}|}}\\ \nonumber &\times& \exp{[-(|\vec{r} - \vec{r}'|/\lambda_{0})]}
 + \vec{\eta}
\end{eqnarray}

{\noindent}Let us first take the case $L \rightarrow \infty $ with
$\lambda_{0}$ finite but large. The dynamic exponent in this case is
$2$, because the kernel effectively renormalises the time scale in a
way that is independent of system size. If we took the two limits
$L\rightarrow \infty$ and $\lambda_0 \rightarrow \infty$ in the
opposite order, as was done in the computer simulations,
the exponential factor would not be present, and the dynamics would be
independent of $L$.  Hence the dynamic exponent would then be $z=0$.

{\it Nature of the long range case:-} The rather curious result of $z =
0$ is obtained for the situation where the screening length
is sent to infinity before taking the thermodynamic limit. Whether the
interaction is considered short or long range depends on with what it
is compared. Physically the short range case describes the situation
where $\lambda_{0}$ is much smaller than the inter-vortex spacing
$\lambda_v$. This is indeed captured in the simulations by setting
$\lambda_{0} = 0$. Physically the long range case describes the
situation where  $L \gg \lambda_{0} \gg \lambda_v$. This is not
captured by setting $\lambda_{0} = \infty$ with $L$ finite.

{\it Reconciliation with the lattice model simulations:-} The critical
dynamics of the lattice simulations and the continuum analysis above do
not apparently agree. We now will show that this is because the
time step in the simulation does not correspond to the physical time
step. The reason is that from the definition of the loop variable
$\vec{l_{i}}$, the net electric field at time $t$ is
${\sc E}^{\alpha}(t) =
\sum_{i}l^{\alpha}_{i}dP[l^{\alpha}_{i}]/dt$.
In the simulations, and
in (4), this has been replaced by
${\sc E}^{\alpha}(t_{MC}) =
\sum_{i}dP[l^{\alpha}_{i}]/dt_{MC}$,
where $t_{MC}$ denotes the Monte
Carlo time and $l^{\alpha}_{i} = \pm 1,0$ only. However in the long
range case and in the short range case at $T_{c}$, where $\xi = L$,
$[l]$ depends on $L$. Hence the physical time is related to the Monte
Carlo time by $t = t_{MC}[l]$ so that the relaxation time is actually
\begin{equation}
\tau \sim L^{3}[l]^{2}/a^{\alpha}_{i}.
\end{equation}

The dynamic exponents for the lattice
model are then $z = 2$ for the short range case and $z = 0$ for the long
range case, in agreement with the analytic calculation based on the
continuum limit equations of motion.  We see that the simulation result
$z_{MC}= 3/2$ or equivalently its corrected form $z=0$ arise from
taking the thermodynamic limit and the long-range of interaction limits
in the incorrect order.  With this correction to the results of the
simulation, the results no longer are consistent with those experiments
reporting $z\sim 1.5$.

{\it Experimental ramifications:-} In experiments performed on bulk
superconductors one would expect the short range limit of the model
above to apply, provided that the interaction range is shorter than the
system size. In such systems, as long as diffusive dynamics for the
vortex degrees of freedom is applicable, a dynamic exponent of $2$ is
predicted by the model above.

What then could be the origin of the behaviour $z \sim 1.5$, if
confirmed, in some experiments? There are at least two possible avenues
for further investigation into the true nature of the critical dynamics
in these systems. The first is to seek experimental evidence for the
existence of hydrodynamic modes which might account for the observed
model E dynamics in transport properties. The 41 meV peak observed in
neutron scattering data is one possible candidate \cite{BGS} although
it does not seem to occur near the origin, while certain
interpretations of the peak-dip-hump structure seen in ARPES are also
suggestive of the existence of a collective mode in the system
\cite{NRM}. A second possibility is to study the crossover from model E
to model A dynamics as the effective coupling of the condensate with
the electromagnetic field tends towards zero (equivalently one can
study the crossover by sending the plasmon gap to zero).

In conclusion, we have explained the simulation results for the
critical dynamics of the superconducting transition in zero field, and
shown that in fact they are consistent with expectations based on the
TDGL.  An extension of this analysis to two dimensions will be
presented elsewhere.

\acknowledgements We thank Jack Lidmar, Mats Wallin, Peter Young and
Steve Teitel for helpful comments on an earlier version of this
manuscript.  We acknowledge support from the National Science
Foundation through grant NSF-DMR-99-70690.


\begin{thebibliography}{99}

\bibitem{Kamal}
S. Kamal, D. Bonn, N. Goldenfeld, P. Hirschfeld, R. Liang and W. N.
Hardy, Phys. Rev. Lett. {\bf 73}, 1845 (1994).

\bibitem{LBH}
R. Liang, D. A. Bonn and W. N. Hardy, Phys Rev. Lett. {\bf 76}, 835
(1996).

\bibitem{SAL1}
M.B. Salamon, J. Shi, N. Overend and M. Howson, Phys. Rev. B {\bf 47},
5520 (1993).

\bibitem{HWS}
M.A. Howson, N. Overend, I.D. Lawrie and M.B. Salamon, Phys. Rev. B
{\bf 51}, 11 984 (1995).

\bibitem{OVR2}
N. Overend, M.A. Howson and I.D. Lawrie, Phys. Rev. Lett. {\bf 72},
3238(1994).

\bibitem{FFH}
D.S. Fisher, M.P.A. Fisher and D.A. Huse, Phys. Rev. B {\bf 43}, 130
(1990).

\bibitem{HH}
P.C. Hohenberg and B.I. Halperin, Rev. Mod. Phys. {\bf 49}, 435 (1977).

\bibitem{NGbook}
See (e.g.) Nigel Goldenfeld, {\it Lectures on Phase Transitions and the
Renormalization Group} (Addison-Wesley, Reading MA, 1992).

\bibitem{JTK}
J.T. Kim, N. Goldenfeld, J. Giapintzakis and D.M. Ginsberg, Phys. Rev B
{\bf 56}, 118 (1997).

\bibitem{HER}
W. Holm, Yu Eltsev and {\"O}. Rapp, Phys. Rev B {\bf 51} 11 992 (1995).

\bibitem{PDR}
A. Pomar et al., Physica C {\bf 218} 257 (1993).

\bibitem{JCB}
J.C. Booth et al,. Phys. Rev. Lett. {\bf 77}, 4438 (1996).

\bibitem{DRS}
R.A. Wickham and A.T. Dorsey, Phys. Rev. B {\bf 61}, 6945 (2000).

\bibitem{MOL}
K. Moloni et al., Phys. Rev. Lett. {\bf 78}, 3173 (1997); ibid. Phys.
Rev. B {\bf 56}, 14784 (1997).

\bibitem{HAN}
S.H. Han, Yu. Eltsev and \"O. Rapp, J. Low Temp. Phys. {\bf 117}, 1259
(1999).

\bibitem{VIL}
J. Villain, J. Phys. (Paris) {\bf 36}, 581 (1977).

\bibitem{LID}
J. Lidmar et al., Phys. Rev. B {\bf 58}, 2827 (1998).

\bibitem{JEN}
H. Weber and H.J. Jensen, Phys. Rev. Lett. {\bf 78}, 2620 (1997).

\bibitem{MIN}
L.M. Jensen, B.J. Kim and P. Minnhagen, Phys. Rev. B {\bf 61}, 15412
(2000); see also, P. Minnhagen, B.J. Kim and H. Weber,
cond-mat/0105323.

\bibitem{KNT}
H. Kleinert, {\it Gauge theories in condensed matter, Vol. 1} (World
Scientific, 1989).

\bibitem{ISI}
See (e.g.) A. Isihara, {\it Statistical Physics} (Academic, 1971).

\bibitem{BGS}
P. Bourges et al., Science {\bf 288}, 1234 (2000).

\bibitem{NRM}
M.R. Norman and H. Ding, Phys. Rev. B {\bf 57}, R11089 (1998).

\end{thebibliography}
\end{document}